\documentclass[aps,prx,twocolumn,superscriptaddress,nofootinbib,floatfix]{revtex4-2}

\usepackage{amsmath,amssymb}
\usepackage{graphicx}
\usepackage{hyperref}
\usepackage{xcolor}

\hypersetup{
  colorlinks=true,
  linkcolor=blue!60!black,
  citecolor=blue!60!black,
  urlcolor=blue!60!black,
}

\newcommand{\bra}[1]{\langle #1 |}
\newcommand{\ket}[1]{| #1 \rangle}
\newcommand{\braket}[2]{\langle #1 | #2 \rangle}
\newcommand{\avg}[1]{\langle #1 \rangle}
\newcommand{\tfd}{\mathrm{TFD}}
\newcommand{\tr}{\mathrm{Tr}}

\begin{document}

\title{No chaos required: traversable wormhole signals survive 98\% coupling deletion}

\author{Sagar Dubey}
\affiliation{Stony Brook University, Stony Brook, NY 11794, USA}
\email{sagar.dubey@stonybrook.edu}

\begin{abstract}
The traversable wormhole protocol in coupled Sachdev-Ye-Kitaev (SYK)
systems produces a transmission signal $C(t)$ widely interpreted as
evidence of holographic dynamics. Recent work has questioned this
interpretation, showing that similar signals arise in generic
thermalizing systems. We address what the signal actually probes by
systematically destroying quantum chaos in the SYK model via random
coupling deletion, while monitoring the transmission signal across the
chaos-to-integrable transition. Using exact diagonalization of the
doubled SYK model at $N=10$ with 50 disorder realizations per sparsity,
supplemented by Krylov-subspace extensions to $N=20$, we find that the
ensemble-averaged peak height varies by less than 1.1\% across a
50-fold sparsification range, even as the underlying spectrum
transitions from Gaussian-unitary-ensemble to sub-Poisson statistics.
A 1{,}200-instance sweep over the inter-system coupling $\mu$ confirms
that the signal is controlled by $\mu$ alone, with no dependence on
internal chaos. We further verify that the thermofield double state
retains its thermal structure under sparsification despite substantial
changes to the state vector, providing a structural explanation for
the invariance. These results indicate that the transmission signal
diagnoses inter-system coupling fidelity rather than holographic
dynamics, and that future quantum-simulation experiments require
independent chaos diagnostics to substantiate gravitational claims.
As a practical consequence, the invariance implies that 98\% of the
Hamiltonian's coupling terms can be discarded (with variance rescaling
of the survivors), reducing the gate count per Trotter step by
approximately $50\times$ at $N=10$ and bringing larger
traversable-wormhole simulations within experimental reach.
\end{abstract}

\maketitle

%=============================================================================
\section{Introduction}
\label{sec:intro}
%=============================================================================

The Sachdev-Ye-Kitaev (SYK) model~\cite{Sachdev:1993,Kitaev:2015} has emerged
as a central tool in the study of quantum gravity in the laboratory. Its
all-to-all random couplings produce maximal chaos~\cite{Maldacena:2015waa},
a solvable large-$N$ limit with an emergent conformal symmetry, and a
holographic dual described by nearly anti-de Sitter (AdS$_2$)
gravity~\cite{Maldacena:2016hyu}. These properties make it a natural candidate
for quantum simulation of gravitational phenomena.

Of particular interest is the traversable wormhole protocol, in which two
copies of the SYK model are coupled by a bilinear interaction
$H_\mathrm{int}$~\cite{Gao:2016bin,Maldacena:2018lmt}. The thermofield double
(TFD) state of the coupled system exhibits a transmission signal $C(t)$
interpreted as information traversing a wormhole connecting two
asymptotically-AdS boundaries. This protocol was realized experimentally on a
quantum processor by Jafferis et al.~\cite{Jafferis:2022uhu}, who reported
observing traversable wormhole dynamics in a heavily sparsified $N=7$ SYK
system.

The interpretation of this experiment has been
debated~\cite{Kobrin:2023cva}. A key question is whether the transmission
signal constitutes evidence for holographic dynamics or merely reflects generic
features of the quantum circuit. Schuster et
al.~\cite{Schuster:2021uvg} demonstrated that the same peaked-size
teleportation mechanism underlying the protocol operates in generic
thermalizing systems, not only in holographic ones. Brown et
al.~\cite{Brown:2019hmk} developed the teleportation-by-size framework that
connects operator growth to traversability. These results suggest the signal
may be more universal than the gravitational interpretation implies.

In this work, we address this question by a systematic numerical study of the
transmission signal across the chaos-to-integrable transition in the
sparsified SYK model. Sparsification---randomly deleting a fraction $1-p$ of
the SYK couplings---drives the model from the fully chaotic GUE regime
through a transition into a non-chaotic, sub-Poisson
regime~\cite{Garcia-Garcia:2020ttf,Xu:2020shn}. If the transmission signal specifically requires the chaotic internal
dynamics of the SYK model---as distinct from merely requiring the
coupling structure---it should degrade as chaos is destroyed. We find
that it does not: the ensemble-averaged signal is invariant under
sparsification, controlled entirely by the inter-system coupling $\mu$.

This result has consequences on three fronts. First, it constrains the
interpretation of quantum gravity simulation experiments: if the
transmission signal is insensitive to whether the system is chaotic or
integrable, it cannot by itself constitute evidence for holographic
dynamics. The signal observed in the Google experiment~\cite{Jafferis:2022uhu}
is consistent with our non-chaotic sparse systems just as well as with a
holographic wormhole, and future experiments will need to supplement it
with independent chaos diagnostics---such as level spacing statistics or
out-of-time-order correlators---to substantiate gravitational claims.
Second, the invariance has an immediate practical payoff: since 98\% of the
Hamiltonian's coupling terms can be discarded without affecting the signal,
the quantum gate count per Trotter step drops by a factor of ${\sim}50$
at $N=10$, with the reduction scaling quartically to ${\sim}200\times$ at
$N=20$ and beyond $1{,}000\times$ at $N=30$. This brings larger traversable
wormhole simulations---currently far beyond the reach of near-term quantum
hardware in their dense form---within experimental feasibility.
Third, together with the results of Schuster et
al.~\cite{Schuster:2021uvg}, our findings establish that the transmission
signal is substantially more universal than its gravitational
interpretation suggests, operating in systems that are neither holographic
nor chaotic.

The remainder of this paper is organized as follows.
Section~\ref{sec:methods} defines the model, the transmission signal, and
the chaos and noise diagnostics. Section~\ref{sec:results} presents the
numerical evidence for signal-chaos decoupling, the $\mu$-control
mechanism, noise-sparsity factorization, disorder fluctuations,
TFD structural robustness, the extension to $N=20$, and the
$N=24$ chaos transition.
Section~\ref{sec:discussion} discusses the structural explanation, the
connection to peaked-size teleportation, gate complexity, and open
questions. Section~\ref{sec:conclusion} summarizes the findings.

%=============================================================================
\section{Model and Methods}
\label{sec:methods}
%=============================================================================

%-----------------------------------------------------------------------------
\subsection{The doubled SYK model}
\label{sec:model}
%-----------------------------------------------------------------------------

We consider $N$ Majorana fermions per side, satisfying
$\{\psi_i, \psi_j\} = 2\delta_{ij}$, with $\psi_i^\dagger = \psi_i$.
These are represented on $N/2$ qubits via the Jordan-Wigner transformation:
\begin{align}
  \psi_{2k-1} &= \Bigl(\prod_{j=1}^{k-1} Z_j\Bigr) X_k\,,
  \label{eq:jw_odd} \\
  \psi_{2k}   &= \Bigl(\prod_{j=1}^{k-1} Z_j\Bigr) Y_k\,,
  \label{eq:jw_even}
\end{align}
for $k = 1, \ldots, N/2$, where $X_k, Y_k, Z_k$ are Pauli matrices on
qubit $k$ and the product is a Jordan-Wigner string that enforces
anticommutation across sites.
The single-side SYK Hamiltonian with $q=4$ is
\begin{equation}
  H_\mathrm{SYK} = -\frac{1}{4!}\sum_{i<j<k<l} J_{ijkl}\,
    \psi_i \psi_j \psi_k \psi_l \,,
\label{eq:syk}
\end{equation}
where the prefactor follows the Kitaev convention $i^{q/2} = i^2 = -1$.
(For $q=4$ the product $\psi_i\psi_j\psi_k\psi_l$ is already Hermitian---reversing
the ordering requires $\binom{4}{2} = 6$ anticommutations, giving
$(-1)^6 = +1$---so this sign is a convention rather than a Hermiticity
requirement; it ensures the standard spectral structure at large $N$.) The couplings are drawn
independently from $J_{ijkl} \sim \mathcal{N}(0, \sigma^2)$ with
\begin{equation}
  \sigma^2 = \frac{6\,J^2}{N^3}\,,
  \label{eq:variance}
\end{equation}
and $J=1$ sets the energy scale. All energies, times, and temperatures are
measured in units of $J$, $1/J$, and $1/J$, respectively.

The doubled system places $2N$ Majorana fermions on a single Jordan-Wigner
chain of $N$ qubits total: qubits $1$ through $N/2$ carry the left
Majoranas ($\psi_0^L, \ldots, \psi_{N-1}^L$), and qubits $N/2+1$ through
$N$ carry the right Majoranas ($\psi_0^R, \ldots, \psi_{N-1}^R$). The full
doubled Hilbert space has dimension $d = 2^N$. The left and right SYK
Hamiltonians, $H_L$ and $H_R$, are built from their respective Majorana
operators using the \emph{same} coupling tensor $J_{ijkl}$---i.e., both
sides share the same disorder realization. Since $H_L$ and $H_R$ act on
disjoint sets of qubits, $[H_L, H_R] = 0$. The two sides are coupled by
the Maldacena-Qi interaction~\cite{Maldacena:2018lmt}:
\begin{equation}
  H_\mathrm{int} = i\mu \sum_{j=0}^{N-1} \psi_j^L \psi_j^R\,,
  \label{eq:hint}
\end{equation}
where the factor of $i$ ensures Hermiticity:
$(i\psi^L\psi^R)^\dagger = -i\psi^R\psi^L = +i\psi^L\psi^R$, using
$\{\psi^L, \psi^R\} = 0$. The full Hamiltonian is
\begin{equation}
  H = H_L + H_R + H_\mathrm{int}\,.
  \label{eq:full_H}
\end{equation}
Throughout this work we use $N=10$ ($d = 2^{10} = 1024$) for the primary
signal and chaos diagnostics, $N=14$ ($d = 2^{14} = 16{,}384$) for level
spacing verification, $N=8$ ($d = 2^8 = 256$) for open-system noise
studies, $N=20$ ($d = 2^{20} = 1{,}048{,}576$) for a larger-$N$
extension using Krylov subspace methods (Sec.~\ref{sec:krylov}), and
$N=24$ ($d = 2^{24} \approx 1.7 \times 10^7$; single-side dimension
$d_s = 4096$) for level spacing analysis across the chaos transition.

%-----------------------------------------------------------------------------
\subsection{Sparsification}
\label{sec:sparsification}
%-----------------------------------------------------------------------------

Following
Refs.~\cite{Garcia-Garcia:2020ttf,Xu:2020shn,Jafferis:2022uhu},
we sparsify the SYK couplings by retaining each $J_{ijkl}$ independently
with probability $p$ and setting it to zero with probability $1-p$. To
preserve ensemble-level statistics, the variance of the surviving couplings
is rescaled:
\begin{equation}
  \avg{J_{ijkl}^2}_\mathrm{sparse} = \frac{6\,J^2}{p\,N^3}\,.
  \label{eq:sparse_variance}
\end{equation}
This ensures that the second moment of the total Hamiltonian is preserved in
the ensemble average, so that bulk spectral properties (bandwidth, mean
energy) remain comparable across sparsities. The same sparsity mask and
rescaling are applied identically to both $H_L$ and $H_R$, so both sides
continue to share the same coupling tensor at every sparsity level. At
$N=10$, the dense model has $\binom{10}{4} = 210$ independent couplings; at
$p = 0.02$, approximately $4$ survive. Crucially, the sparsification acts
only on $H_L$ and $H_R$; the coupling $H_\mathrm{int}$ in
Eq.~\eqref{eq:hint} is unmodified at all sparsity levels. We study nine
sparsity values: $p \in \{1.0,\, 0.5,\, 0.3,\, 0.2,\, 0.1,\, 0.07,\, 0.05,\,
0.03,\, 0.02\}$.

%-----------------------------------------------------------------------------
\subsection{Thermofield double state}
\label{sec:tfd}
%-----------------------------------------------------------------------------

The TFD state at inverse temperature $\beta$ is constructed from the
single-side eigenstates $\{|n\rangle\}$ with eigenvalues $\{E_n\}$:
\begin{equation}
  \ket{\tfd(\beta)} = \frac{1}{\sqrt{Z}} \sum_n e^{-\beta E_n/2}\,
    \ket{n}_L \ket{n}_R\,,
  \label{eq:tfd}
\end{equation}
where $Z = \sum_n e^{-\beta E_n}$ is the partition function.

A subtlety arises in the definition of $|n\rangle_R$. In the abstract
formulation of Refs.~\cite{Maldacena:2018lmt,Maldacena:2013xja}, the TFD
involves a conjugated right eigenstate $|n^*\rangle_R$, with the
conjugation implementing the map $H_R = K H_L K^{-1}$ for an anti-unitary
operator $K$. In our explicit qubit representation, however, the right
Majorana operators $\psi_j^R$ have the \emph{same} matrix structure as their
left counterparts $\psi_j^L$ (both follow the Jordan-Wigner map of
Eqs.~\eqref{eq:jw_odd}--\eqref{eq:jw_even}, applied to their respective
qubits). Consequently, $H_R$ restricted to the right tensor factor is
numerically the same matrix as $H_L$ restricted to the left factor, and no
conjugation is needed: $|n\rangle_R = |n\rangle_L$, with both referring to
the same column of the eigenvector matrix.%
\footnote{In the language of Ref.~\cite{Maldacena:2018lmt}, the
antiunitary operator $K$ acts as complex conjugation in the
computational basis. Since our Jordan-Wigner representation produces
real coupling matrices $J_{ijkl}$ and real Pauli products, the
single-side Hamiltonian $H_L$ is a real symmetric matrix whose
eigenvectors can be chosen real. In this basis $K$ acts trivially on
the right eigenstates, reducing to $|n^*\rangle_R = |n\rangle_R$.}
We verify this convention numerically in Sec.~\ref{sec:tfd_structure}:
the partial trace $\tr_R|\tfd\rangle\langle\tfd|$ reproduces the
canonical thermal state $e^{-\beta H_L}/Z$ to machine precision
($\|\rho_L - \rho_\mathrm{th}\|_F \sim 10^{-16}$) across all
sparsities (Table~\ref{tab:tfd}), confirming that the construction
satisfies the defining TFD property.

In the computational basis, $|n\rangle_L = \sum_a u_{an} |a\rangle_L$ and
$|n\rangle_R = \sum_b u_{bn} |b\rangle_R$, so the TFD is assembled as
\begin{equation}
  \ket{\tfd} = \frac{1}{\sqrt{Z}} \sum_n e^{-\beta E_n/2}
    \sum_{a,b} u_{an}\, u_{bn}\, |a\rangle_L |b\rangle_R \,,
\end{equation}
with the doubled-space index $a \cdot d_s + b$, where $d_s = 2^{N/2}$ is
the single-side dimension and the left index runs slower (standard
Kronecker product ordering). We use $\beta = 8$ throughout, following the parameter regime of
Ref.~\cite{Maldacena:2018lmt}. At this temperature the ground state
Boltzmann weight dominates the TFD ($e^{-\beta E_0/2} \gg
e^{-\beta E_1/2}$ for the typical ground state gap at $N=10$), placing
the system deep in the low-temperature regime where the wormhole signal
is prominent. The fiducial coupling $\mu = 0.1\,J$ is in the weak-coupling
regime ($\mu \ll J$), consistent with the perturbative treatment of
$H_\mathrm{int}$ in the holographic interpretation.

%-----------------------------------------------------------------------------
\subsection{TFD structural diagnostics}
\label{sec:tfd_diagnostics}
%-----------------------------------------------------------------------------

To test whether the TFD retains its thermal character under
sparsification, we compute three diagnostics at each sparsity level
(using $N=10$, $\beta=8$, 30 disorder realizations):
\begin{itemize}
\item \emph{Entanglement entropy}: $S_\mathrm{ent} = -\tr(\rho_L
  \log \rho_L)$, where $\rho_L = \tr_R |\tfd\rangle\langle\tfd|$ is
  the reduced density matrix obtained by tracing out the right
  subsystem. For an exact thermal state,
  $S_\mathrm{ent} = \log Z + \beta \avg{E}$.
\item \emph{Thermal fidelity}: the Frobenius distance $\|\rho_L -
  \rho_\mathrm{th}\|_F$ between the reduced state and the canonical
  thermal state $\rho_\mathrm{th} = e^{-\beta H_L}/Z$.
\item \emph{State overlap}: the absolute overlap
  $|\braket{\tfd_\mathrm{sparse}}{\tfd_\mathrm{dense}}|$ between the
  TFD constructed from the sparsified Hamiltonian and the TFD of the
  dense model (same disorder seed).
\end{itemize}

%-----------------------------------------------------------------------------
\subsection{Krylov subspace methods for $N=20$}
\label{sec:krylov}
%-----------------------------------------------------------------------------

At $N=20$, the doubled Hilbert space has dimension $d = 2^{20} =
1{,}048{,}576$, making explicit matrix storage ($d^2$ complex entries,
${\sim}16$~TB) infeasible. We avoid this by representing $H$ as a
\texttt{LinearOperator} that applies $H_L$, $H_R$, and $H_\mathrm{int}$
via tensor product structure: $H_L$ and $H_R$ are stored as $2^{10}
\times 2^{10}$ matrices and applied to the appropriate tensor factor,
while $H_\mathrm{int}$ uses $N$ rank-one terms. This reduces the operator storage from $O(d^2)$ to $O(d_s^2)$ where
$d_s = 2^{N/2}$; the dominant memory cost is the Krylov basis of $m$
vectors of dimension $d$, requiring $O(md) = O(md_s^2)$.

Time evolution $e^{-iHt}\ket{\psi}$ is computed via the Lanczos
algorithm~\cite{Park:1986}: a Krylov subspace $\mathcal{K}_m =
\mathrm{span}\{|\psi\rangle, H|\psi\rangle, \ldots, H^{m-1}|\psi\rangle\}$
is constructed with $m = 60$, producing an orthonormal basis $V_m$ and
tridiagonal matrix $T_m = V_m^\dagger H V_m$. The time-evolved state is
then $e^{-iHt}|\psi\rangle \approx V_m\, e^{-iT_m t}\, e_1$, where
$e^{-iT_m t}$ is evaluated by diagonalizing the $60 \times 60$ matrix
$T_m$. This avoids full diagonalization of $H$ while achieving
spectral accuracy within the Krylov subspace.

The TFD state is constructed by exact diagonalization of the single-side
Hamiltonian ($d_s = 1024$) and tensor-product assembly. We compute the
transmission signal averaging over $N_\mathrm{sites} = 10$ even-indexed
Majorana sites ($j = 0, 2, 4, \ldots, 18$) rather than all $N=20$
sites, to manage computational cost; by the statistical equivalence
of Majorana sites in the SYK ensemble, this half-set subsampling does
not bias the ensemble average. We use
60 time points over $t \in [0, 30/J]$. The parity-sector level spacing
ratio is computed from the single-side spectrum restricted to the
appropriate fermion parity sector ($N \bmod 8 = 4$ places the $N=20$
model in the GSE universality class~\cite{You:2016}). We use
five sparsity values $p \in \{1.0, 0.5, 0.1, 0.05, 0.02\}$ with 30
disorder realizations each (150 total instances; reduced from 50 at
$N=10$ to manage the increased per-instance cost, and similarly
10 realizations at $N=24$).

%-----------------------------------------------------------------------------
\subsection{Transmission signal}
\label{sec:signal}
%-----------------------------------------------------------------------------

The traversable wormhole transmission signal is defined as
\begin{equation}
  C(t) = \frac{1}{N}\sum_{j=0}^{N-1}
    \bra{\tfd} \psi_j^R(t)\, \psi_j^L(0) \ket{\tfd}\,,
  \label{eq:signal}
\end{equation}
where $\psi_j^R(t) = e^{iHt}\psi_j^R e^{-iHt}$ is the Heisenberg-picture
operator evolved under the full coupled Hamiltonian $H$. We compute $C(t)$
via exact diagonalization of $H$. Writing $H = V\,\mathrm{diag}(E_m)\,V^\dagger$
and inserting two resolutions of the identity in the eigenbasis gives
\begin{equation}
  C_j(t) = \sum_{m,n} A_{mn}^{(j)}\, e^{i(E_m - E_n)t}\,,
  \label{eq:eigen_signal}
\end{equation}
with the spectral weight matrix
\begin{equation}
  A_{mn}^{(j)} = \braket{m}{\tfd}^*\,
    \bra{m}\psi_j^R\ket{n}\,
    \bra{n}\psi_j^L\ket{\tfd}\,,
  \label{eq:Amn}
\end{equation}
where $\ket{m}$, $\ket{n}$ are eigenstates of the full $H$. In practice,
$\braket{m}{\tfd}^*$ and $\braket{n}{\psi_j^L|\tfd}$ are computed as
matrix-vector products in the eigenbasis, and
$\bra{m}\psi_j^R\ket{n} = (V^\dagger \psi_j^R V)_{mn}$ by a single basis
rotation. The Hermiticity of $H$ is enforced exactly before diagonalization
($H \to \tfrac{1}{2}(H + H^\dagger)$) to eliminate floating-point symmetry
breaking.

Time evolution is evaluated on a uniform grid $t \in [0, 30/J]$ with
$120$ points (for $N=10$) or $80$ points (for $N=8$), which is sufficient
to resolve the transmission peak at $t^* \approx 7/J$. From the resulting
time series $|C(t)|$ we extract three observables:
\begin{itemize}
\item \emph{Peak height}: $|C(t^*)|$, the maximum of $|C(t)|$ over the
  time grid.
\item \emph{Peak time}: $t^* = \arg\max_t |C(t)|$.
\item \emph{Full width at half maximum} (FWHM): determined by searching
  left and right from $t^*$ for the first crossing below $|C(t^*)|/2$,
  with linear interpolation between grid points at each half-height
  crossing. If $|C(t)|$ does not drop below $|C(t^*)|/2$ on either side,
  the FWHM is recorded as undefined.
\end{itemize}
All signal computations average over Majorana sites $j = 0, \ldots, N-1$.
For each sparsity value, we perform $50$ independent disorder realizations
(seeds $0$--$49$) and report the ensemble mean and standard error of the
mean (SEM).

%-----------------------------------------------------------------------------
\subsection{Chaos diagnostics}
\label{sec:chaos}
%-----------------------------------------------------------------------------

We diagnose the onset of quantum chaos via the adjacent gap ratio of the
single-side energy spectrum (i.e., the eigenvalues of $H_L$, equivalently
$H_R$, since both share the same coupling tensor):
\begin{equation}
  r_n = \frac{\min(s_n, s_{n+1})}{\max(s_n, s_{n+1})}\,,
  \label{eq:gap_ratio}
\end{equation}
where $s_n = E_{n+1} - E_n$ are consecutive level spacings of the
ordered eigenvalues ($E_1 \leq E_2 \leq \cdots$). The ensemble
average $\avg{r}$ takes the value $\avg{r}_\mathrm{GUE} \approx 0.603$ for
quantum chaotic systems with GUE statistics and
$\avg{r}_\mathrm{Poisson} \approx 0.386$ for integrable (Poisson-distributed)
spectra. At $N=10$ and $N=14$, the SYK model with $q=4$ belongs to GUE
symmetry classes~\cite{You:2016} ($N \bmod 8 = 2$ and $6$,
respectively), making these system sizes appropriate for studying the
chaos-to-integrable transition without complications from additional
symmetry sectors.

%-----------------------------------------------------------------------------
\subsection{Noise model}
\label{sec:noise}
%-----------------------------------------------------------------------------

To study the robustness of the signal under decoherence, we evolve the
density matrix under a Lindblad master equation with single-qubit
dephasing:
\begin{equation}
  \frac{d\rho}{dt} = -i[H,\rho] + \sum_k
    \left(L_k \rho L_k^\dagger
    - \tfrac{1}{2}\{L_k^\dagger L_k, \rho\}\right),
  \label{eq:lindblad}
\end{equation}
with Lindblad operators $L_k = \sqrt{\gamma}\, Z_k$ for each qubit
$k = 1, \ldots, N$, where $\gamma$ is the dephasing rate. For the
dephasing channel ($L_k^\dagger = L_k$, $L_k^2 = \gamma\, I$), the
dissipator reduces to $\gamma\sum_k (Z_k \rho Z_k - \rho)$.

The noisy transmission signal is computed in the Schr\"odinger picture:
\begin{equation}
  C(t) = \frac{1}{N}\sum_j \tr\bigl[\psi_j^R\,\rho_j(t)\bigr]\,,
\end{equation}
where $\rho_j(t)$ evolves under Eq.~\eqref{eq:lindblad} from the initial
state $\rho_j(0) = \psi_j^L |\tfd\rangle\langle\tfd|$, which incorporates
the left Majorana insertion of the transmission protocol.

Numerically, the density matrix is vectorized (flattened in row-major
order to a vector of length $d^2$) and evolved using the explicit
Runge-Kutta method (RK45) from \texttt{scipy.integrate.solve\_ivp}, with
relative tolerance $10^{-8}$ and absolute tolerance $10^{-10}$.
These computations use $N=8$ ($d = 256$, so $d^2 = 65{,}536$ complex
entries per density matrix), which is the largest system size tractable
under the $O(d^2)$ memory and $O(d^3)$ per-step cost of direct Lindblad
evolution. We study six dephasing rates
$\gamma \in \{0, 0.001, 0.003, 0.01, 0.03, 0.1\}$ at four sparsities
$p \in \{1.0, 0.3, 0.1, 0.05\}$, with $30$ disorder realizations each,
using $\mu = 0.1$ and $\beta = 8$ as for the unitary computations.
In a separate computation, we test the joint $(\mu, \gamma)$ dependence
on a $3 \times 4$ grid ($\mu \in \{0.05, 0.1, 0.2\}$,
$\gamma \in \{0, 0.01, 0.03, 0.1\}$) at two sparsities
($p = 1.0$ and $0.1$) with 30 realizations each (720 total instances).

%=============================================================================
\section{Results}
\label{sec:results}
%=============================================================================

%-----------------------------------------------------------------------------
\subsection{Chaos transition under sparsification}
\label{sec:chaos_transition}
%-----------------------------------------------------------------------------

We first establish the chaos-to-integrable transition driven by
sparsification. Figure~\ref{fig:signal_chaos}(a) shows the level spacing
ratio $\avg{r}$ as a function of sparsity $p$ for $N=10$ and $N=14$, each
computed from 50 disorder realizations.

At $N=10$, the spectrum maintains GUE statistics ($\avg{r} \approx 0.59$)
for $p \geq 0.1$. A sharp transition occurs near $p^* \approx 0.07$--$0.10$,
with $\avg{r}$ dropping to $0.46 \pm 0.02$ at $p=0.07$ and reaching
$0.26 \pm 0.04$ at $p=0.02$---well below the Poisson value of 0.386.
(Sub-Poisson statistics indicate level clustering from approximate
degeneracies in the very sparse Hamiltonian, rather than integrability
in the conventional sense; we use ``non-chaotic'' to describe this
regime throughout.)
At $N=14$, the transition shifts to lower sparsity
($p^* \approx 0.02$--$0.03$), consistent with the expectation that more
couplings survive at a given $p$ for larger $N$
($\binom{14}{4} = 1001$ vs.\ $\binom{10}{4} = 210$). At $N=14$, GUE
statistics persist down to $p = 0.05$ ($\avg{r} = 0.597 \pm 0.007$),
with the transition beginning at $p=0.03$ ($\avg{r} = 0.517 \pm 0.016$).
Both system sizes exhibit the same qualitative behavior: dense chaotic
spectra giving way to sub-Poisson statistics at sufficiently low $p$,
with the critical sparsity corresponding to roughly 15--25 surviving
couplings. Table~\ref{tab:level_spacing} presents the full level spacing
data.

\begin{table}[b]
\caption{Level spacing ratio $\avg{r}$ at $N=10$ and $N=14$ (50
realizations per point). Uncertainties are SEM. Classification:
GUE ($\avg{r} \gtrsim 0.55$), transitional ($0.40$--$0.55$),
Poisson-like ($\avg{r} \approx 0.386$),
sub-Poisson ($\avg{r} < 0.386$).}
\label{tab:level_spacing}
\begin{ruledtabular}
\begin{tabular}{lcccc}
$p$ & \multicolumn{2}{c}{$N=10$} & \multicolumn{2}{c}{$N=14$} \\
    & $\avg{r}$ & Class & $\avg{r}$ & Class \\
\hline
1.000 & $0.589 \pm 0.009$ & GUE & $0.601 \pm 0.005$ & GUE \\
0.500 & $0.587 \pm 0.011$ & GUE & $0.591 \pm 0.005$ & GUE \\
0.300 & $0.602 \pm 0.011$ & GUE & $0.606 \pm 0.005$ & GUE \\
0.200 & $0.594 \pm 0.009$ & GUE & $0.604 \pm 0.005$ & GUE \\
0.100 & $0.568 \pm 0.010$ & GUE & $0.591 \pm 0.005$ & GUE \\
0.070 & $0.460 \pm 0.020$ & Trans. & $0.597 \pm 0.006$ & GUE \\
0.050 & $0.394 \pm 0.028$ & P.-like & $0.597 \pm 0.007$ & GUE \\
0.030 & $0.324 \pm 0.032$ & Sub-P. & $0.517 \pm 0.016$ & Trans. \\
0.020 & $0.256 \pm 0.037$ & Sub-P. & $0.422 \pm 0.022$ & Trans. \\
\end{tabular}
\end{ruledtabular}
\end{table}

To extend the chaos transition analysis to a third system size, we
computed the parity-sector level spacing ratio at $N=24$
($\binom{24}{4} = 10{,}626$ total couplings, 70 instances). At this
system size, $N \bmod 8 = 0$ places the model in the GOE universality
class~\cite{You:2016}, with expected
$\avg{r}_\mathrm{parity} \approx 0.536$. The results, summarized in
Table~\ref{tab:n24_level_spacing}, reveal a sharp chaos transition
between $p = 0.003$ ($\avg{r}_\mathrm{parity} = 0.515 \pm 0.011$,
${\sim}33$ surviving couplings) and $p = 0.002$
($\avg{r}_\mathrm{parity} = 0.333 \pm 0.058$, ${\sim}21$ couplings).
For $p \geq 0.005$, the spectrum maintains GOE statistics
($\avg{r}_\mathrm{parity} = 0.528$--$0.532$, slightly below the
asymptotic GOE value of $0.536$ due to finite-size effects at
$d_s = 4096$ and limited realization count), while at $p \leq 0.002$
it falls well below the Poisson value of $0.386$. The critical coupling
count of ${\sim}21$--$33$ is consistent with the threshold of
${\sim}15$--$25$ identified at $N=10$ and $N=14$, providing a third
confirmation point for the hypothesis that random-matrix universality
requires a roughly \emph{constant} number of coupling terms,
independent of the total number available. The full transmission signal
computation at $N=24$ was not feasible with available resources (see
Sec.~\ref{sec:outlook}), so the signal-chaos decoupling cannot be
directly verified at this system size.

\begin{table}[b]
\caption{Parity-sector level spacing ratio $\avg{r}_\mathrm{parity}$
at $N=24$ (10 realizations per point, GOE universality class,
expected $\avg{r} \approx 0.536$). $n_\mathrm{coup}$: mean number of
surviving couplings. The chaos transition occurs between $p=0.003$
and $p=0.002$, consistent with the universal threshold of
${\sim}15$--$25$ surviving couplings found at smaller $N$.}
\label{tab:n24_level_spacing}
\begin{ruledtabular}
\begin{tabular}{lccc}
$p$ & $\avg{r}_\mathrm{par}$ & $n_\mathrm{coup}$ & Class \\
\hline
0.500 & $0.532 \pm 0.002$ & 5292 & GOE \\
0.050 & $0.532 \pm 0.001$ & 528 & GOE \\
0.010 & $0.528 \pm 0.001$ & 105 & GOE \\
0.005 & $0.529 \pm 0.001$ & 52 & GOE \\
0.003 & $0.515 \pm 0.011$ & 33 & Trans. \\
0.002 & $0.333 \pm 0.058$ & 21 & Sub-P. \\
0.001 & $0.306 \pm 0.058$ & 11 & Sub-P. \\
\end{tabular}
\end{ruledtabular}
\end{table}

%-----------------------------------------------------------------------------
\subsection{Signal invariance across the chaos transition}
\label{sec:decoupling}
%-----------------------------------------------------------------------------

\begin{figure}[t]
  \centering
  \includegraphics[width=\columnwidth]{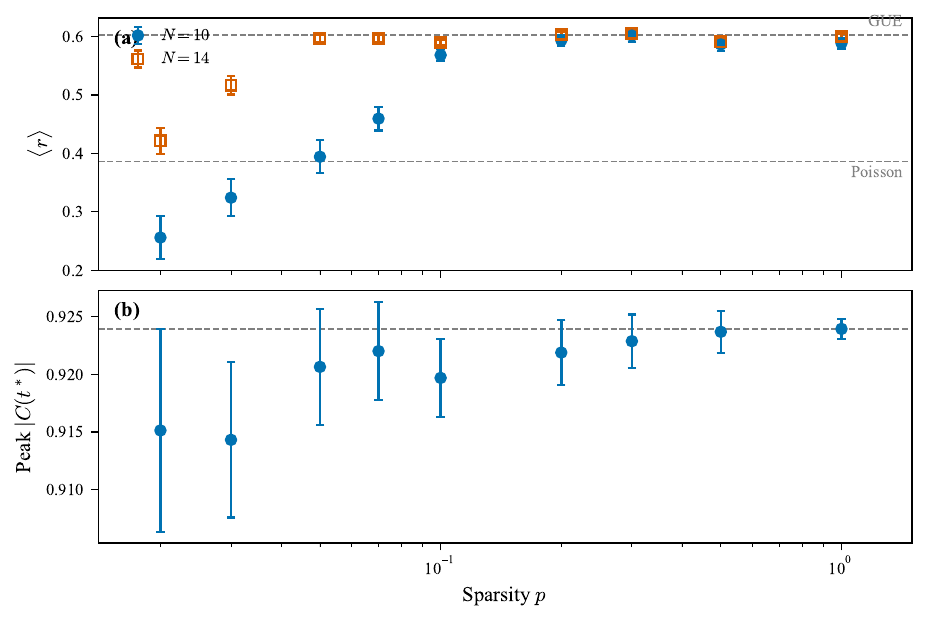}
  \caption{Signal-chaos decoupling at $N=10$, $\beta=8$, $\mu=0.1$
  (50 realizations per point). \textbf{(a)} Level spacing ratio $\avg{r}$
  versus sparsity $p$ for $N=10$ (filled circles) and $N=14$ (open squares).
  Dashed lines indicate GUE ($\avg{r} = 0.603$) and Poisson
  ($\avg{r} = 0.386$) reference values. \textbf{(b)} Transmission peak
  height $|C(t^*)|$ versus $p$. Error bars denote SEM. The dashed line
  marks the dense ($p=1.0$) mean. The signal varies by less than 1.1\%
  despite a complete loss of quantum chaos.}
  \label{fig:signal_chaos}
\end{figure}

Figure~\ref{fig:signal_chaos}(b) shows the main result: the
transmission peak height $|C(t^*)|$ as a function of sparsity, computed
from 50 disorder realizations at each of nine sparsity values from
$p=1.0$ to $p=0.02$. The peak height is $|C(t^*)| = 0.924 \pm 0.001$
at $p=1.0$ and $0.915 \pm 0.009$ at $p=0.02$. The maximum deviation
from the dense mean across all sparsities is less than 1.1\%, and all
pairwise $z$-scores are below 1.4---far from statistical significance.
The peak time is equally invariant: $t^* = 7.27$--$7.34$ across the
full sparsity range, with SEM below $0.06/J$ at every point. The FWHM
of the transmission peak is also sparsity-independent within
uncertainties: $\mathrm{FWHM} \approx 17$--$20\,J^{-1}$ across all nine sparsities,
with no systematic trend.

This invariance is striking when contrasted with the level spacing data
in Fig.~\ref{fig:signal_chaos}(a): the spectrum transitions from fully
chaotic to deeply non-chaotic, yet the transmission signal is unchanged.
The standard deviation of the peak height across realizations does grow
substantially, from 0.006 at $p=1.0$ to 0.061 at $p=0.02$ (a factor
of ${\sim}10$), as shown in Fig.~\ref{fig:disorder}. The invariance
is therefore a property of the ensemble mean, not of individual disorder
realizations.

Table~\ref{tab:transmission} summarizes the numerical data.

\begin{table}[b]
\caption{Transmission peak statistics at $N=10$, $\beta=8$, $\mu=0.1$
(50 realizations). SEM denotes the standard error of the mean; Std is
the standard deviation across realizations.}
\label{tab:transmission}
\begin{ruledtabular}
\begin{tabular}{lccccc}
$p$ & $\avg{r}$ & Mean $|C(t^*)|$ & SEM & Std & Ratio \\
\hline
1.000 & 0.589 & 0.9239 & 0.0009 & 0.006 & 1.000 \\
0.500 & 0.587 & 0.9237 & 0.0018 & 0.013 & 1.000 \\
0.300 & 0.602 & 0.9229 & 0.0023 & 0.016 & 0.999 \\
0.200 & 0.594 & 0.9219 & 0.0028 & 0.020 & 0.998 \\
0.100 & 0.568 & 0.9197 & 0.0033 & 0.024 & 0.995 \\
0.070 & 0.460 & 0.9220 & 0.0042 & 0.030 & 0.998 \\
0.050 & 0.394 & 0.9207 & 0.0050 & 0.035 & 0.997 \\
0.030 & 0.324 & 0.9143 & 0.0067 & 0.047 & 0.990 \\
0.020 & 0.256 & 0.9151 & 0.0087 & 0.061 & 0.990 \\
\end{tabular}
\end{ruledtabular}
\end{table}

%-----------------------------------------------------------------------------
\subsection{The coupling $\mu$ controls the signal}
\label{sec:mu}
%-----------------------------------------------------------------------------

\begin{figure}[t]
  \centering
  \includegraphics[width=\columnwidth]{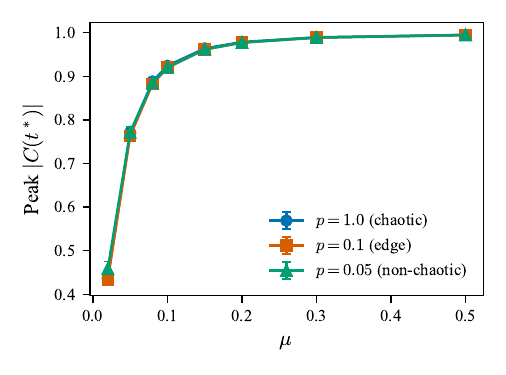}
  \caption{Transmission peak height versus inter-system coupling $\mu$
  for three sparsity levels: $p=1.0$ (chaotic, circles), $p=0.1$ (edge
  of chaos, squares), and $p=0.05$ (non-chaotic, triangles). $N=10$,
  $\beta=8$, 50 realizations. Error bars denote SEM. The three curves
  overlap at each $\mu$ value, confirming that the signal is controlled
  by the coupling, not by the internal dynamics.}
  \label{fig:mu_mechanism}
\end{figure}

To identify the control parameter, we perform a sweep over the coupling
strength $\mu$ at three representative sparsities: $p = 1.0$ (fully chaotic),
$p = 0.1$ (edge of chaos), and $p = 0.05$ (non-chaotic). We compute the
transmission peak at eight coupling values $\mu \in \{0.02, 0.05, 0.08,
0.10, 0.15, 0.20, 0.30, 0.50\}$, with 50 disorder realizations each
(1{,}200 total instances).

Figure~\ref{fig:mu_mechanism} shows the result: the peak height varies
strongly with $\mu$, from $|C(t^*)| = 0.44 \pm 0.02$ at $\mu = 0.02$ to
$|C(t^*)| = 0.995 \pm 0.0002$ at $\mu = 0.5$. However, at each $\mu$
value, the three sparsity curves are statistically indistinguishable. The
maximum deviation between any two sparsities is $3.5\%$ at
$\mu = 0.02$, where the signal is weakest and disorder fluctuations
are largest; pairwise $z$-scores at this point are 0.2, 1.3, and 1.3,
all well below statistical significance. The deviation drops below
$0.7\%$ for $\mu \geq 0.05$, and at the fiducial value $\mu = 0.1$
the three sparsities agree to within $0.27\%$.

The same pattern holds for the FWHM (computed from a separate
higher-resolution time grid with 200 points over $t \in [0, 50/J]$):
the FWHM varies from $29.1 \pm 0.2$ at $\mu = 0.02$ to
$2.090 \pm 0.003$ at $\mu = 0.5$, but shows no dependence on sparsity,
with inter-sparsity deviations below $2.1\%$ at every $\mu$ value and
below $0.4\%$ for $\mu \geq 0.05$. These results confirm that the
transmission signal is a function of $\mu$ alone:
$|C(t^*)| = f(\mu)$, with no dependence on $p$.

%-----------------------------------------------------------------------------
\subsection{Noise-sparsity factorization}
\label{sec:noise_results}
%-----------------------------------------------------------------------------

\begin{figure}[t]
  \centering
  \includegraphics[width=\columnwidth]{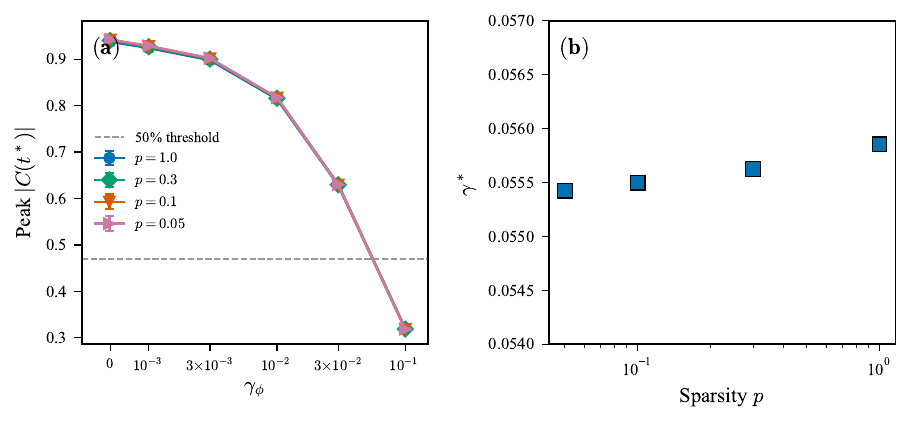}
  \caption{Noise robustness of the transmission signal at $N=8$, $\beta=8$,
  $\mu=0.1$ (30 realizations). \textbf{(a)} Peak height $|C(t^*)|$ versus
  dephasing rate $\gamma$ for four sparsity levels. The curves overlap,
  confirming that noise sensitivity is independent of sparsity. The dashed
  line marks the 50\% degradation threshold. \textbf{(b)} Critical dephasing
  rate $\gamma^*$ (at which the signal drops to 50\% of its noiseless value)
  versus sparsity. $\gamma^*$ is constant at $\approx 0.055\,J$.}
  \label{fig:noise}
\end{figure}

We investigate whether the noise robustness of the signal also decouples
from the internal dynamics. Figure~\ref{fig:noise}(a) shows the peak height
as a function of dephasing rate $\gamma$ at four sparsity levels ($p = 1.0,
0.3, 0.1, 0.05$), computed with $N=8$ and 30 realizations per point.

The noiseless $N=8$ peak height ($|C(t^*)| \approx 0.94$) is slightly
higher than the $N=10$ value ($\approx 0.924$), a finite-size effect
whose precise origin we do not analyze here. The relevant quantity for
the noise analysis is not the absolute peak height but the fractional
degradation as a function of $\gamma$, which is well-defined at any $N$.

We note that the chaos transition at $N=8$ occurs at higher sparsity
than at $N=10$, since $\binom{8}{4} = 70$ (vs.\ 210 at $N=10$). We
computed the level spacing ratio at $N=8$ for the four noise sparsities
and find $\avg{r} = 0.447 \pm 0.013$ at $p=1.0$ and
$\avg{r} = 0.344 \pm 0.041$ at $p=0.05$. (At $N=8$, $N\!\bmod 8 = 0$ places the model in the GOE
symmetry class~\cite{You:2016}. The full-spectrum $\avg{r}$
falls below the single-sector GOE value of $0.536$ because the
computation includes levels from both fermion parity sectors;
inter-sector spacings lack the level repulsion present within each
sector, suppressing $\avg{r}$ toward Poisson-like values.) The noise
sparsity range thus spans a regime with changing spectral statistics,
confirming that the noise-sparsity independence is not an artifact of
all four sparsities being in the same dynamical regime.

The curves for all four sparsities overlap at each noise level, with
deviations below 0.5\%. The signal degrades smoothly from
$|C(t^*)| \approx 0.94$ at $\gamma = 0$ to $|C(t^*)| \approx 0.32$ at
$\gamma = 0.1$, following the same trajectory regardless of sparsity.
Figure~\ref{fig:noise}(b) shows the critical dephasing rate $\gamma^*$,
defined as the rate at which the signal falls to 50\% of its noiseless
value. We find $\gamma^* \approx 0.0554$--$0.0559\,J$, varying by less
than 1\% across sparsities.

Combined with the $\mu$-sweep results of Sec.~\ref{sec:mu}, these data
show that the signal is independent of sparsity at every tested
$(\mu, \gamma)$ pair. To test this jointly, we computed the peak height
on a $3 \times 4$ grid of $(\mu, \gamma)$ values
($\mu \in \{0.05, 0.1, 0.2\}$, $\gamma \in \{0, 0.01, 0.03, 0.1\}$)
at two sparsities ($p = 1.0$ and $0.1$) with 30 realizations each.
At every grid point, the two sparsities agree within $2.1\%$, with
all pairwise $z$-scores below 1.1. These data are consistent with
a factorized form $|C(t^*)| \approx f(\mu) \cdot g(\gamma)$ with
no $p$ dependence. A singular value decomposition of the
$3 \times 4$ mean peak matrix yields a rank-1 residual of $8.6\%$
($\sigma_2/\sigma_1 = 0.086$), indicating that the $\mu$ and $\gamma$
effects are approximately but not exactly multiplicatively separable.
The dominant interaction is at low $\mu$ and high $\gamma$, where the
signal is weakest and the relative corrections are largest.
The sparsity independence, however, holds at each grid point
regardless of the factorization structure.

%-----------------------------------------------------------------------------
\subsection{Disorder fluctuations}
\label{sec:disorder}
%-----------------------------------------------------------------------------

\begin{figure}[t]
  \centering
  \includegraphics[width=\columnwidth]{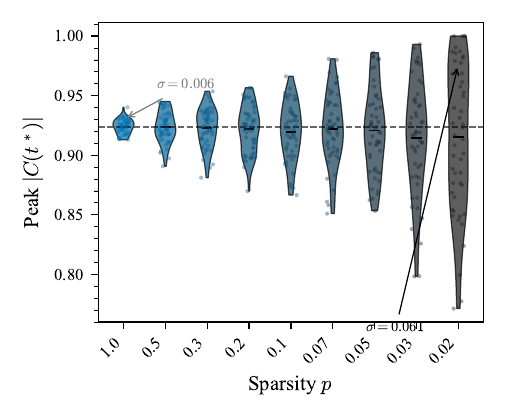}
  \caption{Distribution of peak heights across 50 disorder realizations at
  each sparsity (violin plot with jittered individual data points). The mean
  (horizontal markers inside each violin) is approximately constant across
  all sparsities, but the spread grows from
  $\sigma = 0.006$ at $p=1.0$ to $\sigma = 0.061$ at $p=0.02$. Dashed
  line: dense mean.}
  \label{fig:disorder}
\end{figure}

While the ensemble mean is invariant, the variance across disorder
realizations grows substantially at low sparsity
(Fig.~\ref{fig:disorder}). At $p=1.0$, the standard deviation of the peak
height across 50 realizations is $\sigma = 0.006$, corresponding to
relative fluctuations of 0.7\%. At $p=0.02$, $\sigma = 0.061$ (6.7\%
relative fluctuations)---a tenfold increase. The scaling is approximately
$\sigma \propto p^{-1/2}$: a power-law fit to $\log\sigma$ vs.\ $\log p$
gives an exponent of $-0.53$, close to the $-1/2$ expected from
central-limit-theorem arguments when $p\binom{N}{4}$ independent terms
contribute to the Hamiltonian.

This has practical implications for experiments, which probe a single
Hamiltonian instance rather than an ensemble average. While the expected
signal is preserved at all sparsities, a single realization at very low $p$
may deviate from the mean by several percent. At $p = 0.1$ (${\sim}21$
surviving couplings), single-instance fluctuations are ${\sim}2.6\%$,
comparable to typical experimental uncertainties. At $p = 0.02$
(${\sim}4$ couplings), fluctuations reach ${\sim}6.7\%$, and a single
realization may deviate from the ensemble mean by $2\sigma \approx 13\%$.
Experimentalists should therefore either average over multiple
Hamiltonian instances or choose a sparsity level ($p \gtrsim 0.1$) where
single-instance fluctuations remain manageable.

%-----------------------------------------------------------------------------
\subsection{TFD structural robustness}
\label{sec:tfd_structure}
%-----------------------------------------------------------------------------

\begin{figure}[t]
  \centering
  \includegraphics[width=\columnwidth]{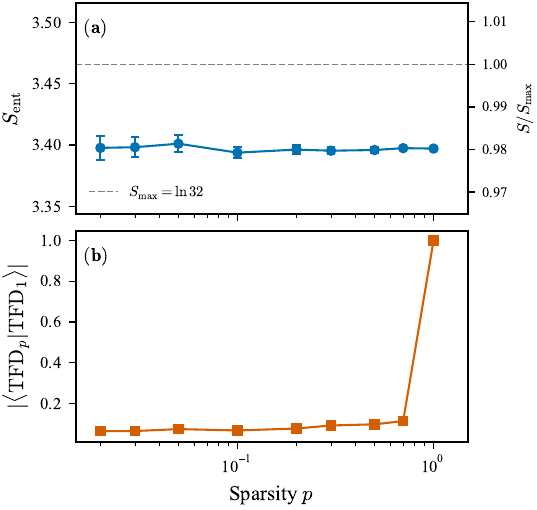}
  \caption{TFD structural diagnostics across the sparsity range at
  $N=10$, $\beta=8$ (30 realizations). \textbf{(a)} Entanglement
  entropy $S_\mathrm{ent}$ of the reduced state $\rho_L$. The entropy
  is nearly constant ($<0.2\%$ variation), close to the maximal value
  $S_\mathrm{max} = \ln 32$. \textbf{(b)} Overlap of the sparse TFD
  with the dense ($p=1$) TFD. The state vector changes dramatically
  (overlap drops below 0.11 at $p \leq 0.5$), yet the thermal
  properties are preserved.}
  \label{fig:tfd_structure}
\end{figure}

The signal invariance raises a structural question: does the TFD state
itself change under sparsification, or is the signal invariant because
the initial state is unchanged? Figure~\ref{fig:tfd_structure} and
Table~\ref{tab:tfd} present the TFD
structural diagnostics defined in Sec.~\ref{sec:tfd_diagnostics},
computed at $N=10$, $\beta=8$, $\mu=0.1$ with 30 disorder
realizations.

\begin{table}[b]
\caption{TFD structural diagnostics at $N=10$, $\beta=8$ (30
realizations). $S_\mathrm{ent}$: entanglement entropy
($S_\mathrm{max} = \ln 32 = 3.47$).
$\|\rho_L - \rho_\mathrm{th}\|_F$: thermal fidelity error.
$|\langle\mathrm{TFD}_p|\mathrm{TFD}_1\rangle|$: overlap with
dense TFD. Uncertainties are SEM.}
\label{tab:tfd}
\begin{ruledtabular}
\begin{tabular}{lcccc}
$p$ & $S_\mathrm{ent}$ & $S/S_\mathrm{max}$ & Thermal err. & Overlap \\
\hline
1.000 & $3.397 \pm 0.001$ & 0.980 & $2 \times 10^{-16}$ & 1.000 \\
0.500 & $3.396 \pm 0.002$ & 0.980 & $2 \times 10^{-16}$ & $0.097$ \\
0.300 & $3.395 \pm 0.003$ & 0.980 & $2 \times 10^{-16}$ & $0.091$ \\
0.100 & $3.394 \pm 0.005$ & 0.979 & $2 \times 10^{-16}$ & $0.067$ \\
0.050 & $3.401 \pm 0.007$ & 0.981 & $2 \times 10^{-16}$ & $0.073$ \\
0.020 & $3.398 \pm 0.010$ & 0.980 & $2 \times 10^{-16}$ & $0.063$ \\
\end{tabular}
\end{ruledtabular}
\end{table}

The entanglement entropy $S_\mathrm{ent}$ is strikingly invariant:
it varies by less than $0.21\%$ across the full sparsity range
($3.394$ at $p=0.1$ to $3.401$ at $p=0.05$), and remains at
$97.9$--$98.1\%$ of the maximum value $S_\mathrm{max} = \ln 32
\approx 3.47$ at every sparsity. The thermal fidelity error is at
machine precision (${\sim}10^{-16}$) for all sparsities, confirming
that $\rho_L$ is an exact Gibbs state of $H_L$ regardless of
sparsification. This follows from the construction of the TFD via
exact diagonalization [Eq.~\eqref{eq:tfd}], but serves as a
numerical consistency check: the TFD at every sparsity is a
legitimate thermal state of its respective Hamiltonian.

The state overlap tells a different story. At $p = 0.5$, the TFD of
the sparse model has only ${\sim}9.7\%$ overlap with the dense TFD
(same disorder seed), dropping to ${\sim}6.3\%$ at $p = 0.02$. The
individual eigenstates of $H_{L,R}$ change drastically under
sparsification---as expected, since the Hamiltonian matrix changes
substantially---yet the thermodynamic structure of the TFD, as
measured by $S_\mathrm{ent}$ and $\rho_\mathrm{th}$, is preserved.

This resolves the structural question: the signal is invariant not
because the initial state is unchanged, but because the
\emph{thermal properties} of the initial state (entanglement
entropy, reduced density matrix) are preserved by the variance
rescaling, while the state itself changes significantly.

%-----------------------------------------------------------------------------
\subsection{Extension to $N=20$}
\label{sec:n20}
%-----------------------------------------------------------------------------

To test whether the signal-chaos decoupling persists at larger system
sizes, we extend the analysis to $N=20$ Majorana fermions per side using
the Krylov subspace methods of Sec.~\ref{sec:krylov}. At $N=20$, the
doubled Hilbert space has dimension $d = 2^{20} \approx 10^6$, and the
dense model has $\binom{20}{4} = 4{,}845$ independent couplings---a
factor of 23 more than at $N=10$. Even at the most extreme sparsity
tested ($p=0.02$), approximately 97 couplings survive, well above the
${\sim}15$--25 threshold for the onset of quantum chaos identified in
Sec.~\ref{sec:chaos_transition}. Consequently, the $N=20$ system remains
in the chaotic regime (GSE universality class, $\avg{r}_\mathrm{parity}
\approx 0.676$) at all tested sparsities.

Table~\ref{tab:n20_transmission} presents the results of the $N=20$
sweep. The ensemble-averaged peak height varies by less than
$0.19\%$ across the full sparsity range, from $|C(t^*)| = 0.9029 \pm
0.0003$ at $p=1.0$ to $0.9031 \pm 0.0023$ at $p=0.02$. All pairwise
$z$-scores relative to the dense mean are below 1.1. The peak time $t^* \approx 7.12\,J^{-1}$ and FWHM
$\approx 10.05\,J^{-1}$ are consistent across all sparsities to
within statistical uncertainty.

\begin{table}[b]
\caption{$N=20$ transmission peak statistics ($\beta=8$, $\mu=0.1$,
30 realizations). Krylov dimension $m=60$, 10 Majorana sites averaged.
The parity-sector level spacing ratio $\avg{r}_\mathrm{parity}$ is
consistent with GSE ($\approx 0.676$) at all sparsities.}
\label{tab:n20_transmission}
\begin{ruledtabular}
\begin{tabular}{lccccc}
$p$ & $\avg{r}_\mathrm{par}$ & Mean $|C(t^*)|$ & SEM & Std & Dev. \\
\hline
1.000 & $0.677 \pm 0.002$ & 0.9029 & 0.0003 & 0.0016 & --- \\
0.500 & $0.672 \pm 0.002$ & 0.9027 & 0.0007 & 0.0037 & 0.03\% \\
0.100 & $0.677 \pm 0.002$ & 0.9026 & 0.0010 & 0.0057 & 0.04\% \\
0.050 & $0.673 \pm 0.002$ & 0.9012 & 0.0016 & 0.0090 & 0.19\% \\
0.020 & $0.674 \pm 0.002$ & 0.9031 & 0.0023 & 0.0126 & 0.02\% \\
\end{tabular}
\end{ruledtabular}
\end{table}

The $N=20$ data confirms two predictions from the $N=10$ analysis.
First, the signal invariance persists at a system size where the
Hilbert space is three orders of magnitude larger. Second, the
disorder fluctuations grow as $\sigma \propto p^{-1/2}$ (the
standard deviation increases from 0.0016 at $p=1.0$ to 0.0126 at
$p=0.02$, a ratio of 7.9), consistent with the central-limit-theorem
scaling found at $N=10$.

The crucial difference from the $N=10$ case is that the chaos
transition lies below the tested sparsity range: with 4{,}845 total
couplings, $p=0.02$ retains ${\sim}97$ couplings---far above the chaos
threshold. The $N=20$ data therefore confirms signal invariance
\emph{within} the chaotic regime over a $50\times$ sparsification
range, but does not probe the non-chaotic regime. Reaching the chaos
threshold at $N=20$ would require $p^* \sim 25/4845 \approx 0.005$,
corresponding to a $200\times$ gate reduction. The structural argument
of Sec.~\ref{sec:mechanism} predicts the decoupling will persist
through this transition, but direct verification remains a target for
future work.

%=============================================================================
\section{Discussion}
\label{sec:discussion}
%=============================================================================

%-----------------------------------------------------------------------------
\subsection{Why the signal is coupling-controlled}
\label{sec:mechanism}
%-----------------------------------------------------------------------------

The signal-chaos decoupling has a simple structural explanation. The
Hamiltonian $H = H_L + H_R + H_\mathrm{int}$ separates into components
with distinct physical roles: $H_L$ and $H_R$ govern the internal
dynamics (chaos, scrambling, thermalization) of each subsystem, while
$H_\mathrm{int}$ controls information transfer between them.

The transmission signal $C(t)$ [Eq.~\eqref{eq:signal}] measures precisely
this inter-system information transfer. Sparsification modifies $H_L$ and
$H_R$ but leaves $H_\mathrm{int}$ untouched. The variance rescaling
[Eq.~\eqref{eq:sparse_variance}] further ensures that ensemble-level
thermodynamic properties of each side are preserved, so the TFD state
retains the same average thermal structure. The signal thus probes a sector
of the Hamiltonian that is invariant under sparsification, explaining the
observed decoupling.

This argument is supported quantitatively by the TFD structural
analysis of Sec.~\ref{sec:tfd_structure}. Although the TFD state
vector changes substantially under sparsification (overlap with the
dense TFD drops below $0.10$ at $p = 0.5$), the thermal properties
that enter the signal---the entanglement entropy, the reduced density
matrix, and the partition function---are preserved to better than
$0.2\%$. The initial state carries information about the internal
dynamics through its eigenstates, but the variance rescaling
[Eq.~\eqref{eq:sparse_variance}] ensures that the thermal structure
relevant to the signal is invariant.

%-----------------------------------------------------------------------------
\subsection{Implications for quantum gravity experiments}
\label{sec:experiments}
%-----------------------------------------------------------------------------

Our results bear directly on the interpretation of quantum gravity
simulation experiments. The experiment of Jafferis et
al.~\cite{Jafferis:2022uhu} used a heavily sparsified $N=7$ SYK system
with approximately 5 surviving couplings---a regime where, based on our
findings, the system is below the chaos threshold. Our analysis shows that
the transmission signal would be present at this sparsity regardless of
whether the dynamics are chaotic, because the signal probes the coupling
$H_\mathrm{int}$, not the internal dynamics. This is consistent with the
critique of Kobrin et al.~\cite{Kobrin:2023cva}, who argued that the
observed signal does not constitute evidence of holographic dynamics.

More broadly, our findings suggest that future traversable wormhole
experiments should report chaos diagnostics---such as the level spacing
ratio $\avg{r}$, the spectral form factor, or out-of-time-order
correlators (OTOCs)---alongside the transmission signal. The signal alone is
not sufficient evidence for holographic dynamics; it confirms that the
inter-system coupling is functional but does not probe the internal
dynamics.

We can map out which observables depend on the internal dynamics and
which do not:
\begin{itemize}
  \item \textit{Chaos-independent}: transmission peak height, peak time,
    FWHM, noise threshold $\gamma^*$, TFD entanglement entropy.
  \item \textit{Chaos-dependent}: level spacing ratio, spectral form
    factor, OTOC growth rate, TFD state overlap with dense.
\end{itemize}
The first category probes $H_\mathrm{int}$; the second probes $H_{L,R}$.
A complete experimental demonstration of holographic dynamics should
include observables from both categories.

%-----------------------------------------------------------------------------
\subsection{Connection to peaked-size teleportation}
\label{sec:teleportation}
%-----------------------------------------------------------------------------

Our findings complement the results of Schuster et
al.~\cite{Schuster:2021uvg}, who showed that the quantum teleportation
mechanism underlying the traversable wormhole protocol---peaked-size
teleportation---operates in generic thermalizing systems, not only in
holographic ones. Where their work demonstrated that \textit{thermalization}
is sufficient (without holography), we show that the signal persists even
\textit{without random-matrix universality}: at $p=0.02$, the system has
sub-Poisson level statistics, yet the transmission signal is unchanged. Together, these results establish that the traversable
wormhole signal is substantially more universal than the gravitational
interpretation suggests.

The teleportation-by-size framework of Brown et al.~\cite{Brown:2019hmk}
provides further context. The key requirement for the protocol is not
maximal chaos or holographic scrambling, but rather that the operator
$\psi_j^R(t)$ develops a peaked size distribution at the appropriate time.
Our results suggest this peaked-size condition is met across the full
sparsity range, a consequence of the TFD state structure and the coupling
$H_\mathrm{int}$ rather than properties of $H_{L,R}$.

%-----------------------------------------------------------------------------
\subsection{Gate complexity reduction}
\label{sec:gates}
%-----------------------------------------------------------------------------

The signal-chaos decoupling has an immediate practical consequence for quantum
simulation. The dense SYK Hamiltonian at $q=4$ has $\binom{N}{4}$ coupling
terms, each requiring $O(N)$ gates after Jordan-Wigner transformation,
giving $O(N^5)$ gates per Trotter step. At sparsity $p$, only
$p\binom{N}{4}$ terms survive, reducing the gate count to $O(pN^5)$.

Since the signal is invariant down to $p=0.02$, 98\% of the coupling terms
can be discarded without affecting the signal. At $N=10$, the dense
Hamiltonian requires ${\sim}1{,}260$ controlled-NOT (CNOT) gates per Trotter
step, while $p=0.02$ requires ${\sim}24$---a $52\times$ reduction. Because
the chaos threshold requires a roughly constant number of couplings
(${\sim}15$--$25$) while the total $\binom{N}{4}$ grows as $N^4/24$, the gate
reduction factor scales quartically with system size: at $N=20$ the
estimated reduction exceeds $200\times$, and at $N=30$ it exceeds
$1{,}000\times$. For context, the Google Sycamore traversable wormhole
experiment~\cite{Jafferis:2022uhu} used a heavily compressed $N=10$
Hamiltonian with only 5 coupling terms; our results show that such
aggressive simplification preserves the signal exactly, and that this
strategy extends systematically to larger system sizes where the full
Hamiltonian is far beyond the reach of current hardware.

We note that the variance rescaling [Eq.~\eqref{eq:sparse_variance}]
increases the spectral norm of individual coupling terms by $1/\sqrt{p}$,
which may require smaller Trotter steps to maintain a fixed simulation
error, partially offsetting the per-step savings. The net gate reduction
after accounting for Trotterization error bounds remains substantial but
is smaller than the naive per-step estimate.

%-----------------------------------------------------------------------------
\subsection{Outlook: scope, extensions, and the question of sufficiency}
\label{sec:outlook}
%-----------------------------------------------------------------------------

The signal-chaos decoupling established here is a numerical result at
system sizes up to $N=20$ ($d = 2^{20} \approx 10^6$). The Krylov
extension of Sec.~\ref{sec:n20} confirms the signal invariance at
$N=20$ within the chaotic regime (all tested sparsities retain GSE
statistics), but the non-chaotic window shrinks with $N$: the critical
sparsity scales as $p^* \sim N^{-4}$ (since chaos requires a roughly
constant number of ${\sim}15$--25 couplings while $\binom{N}{4} \sim
N^4/24$ grows quartically). At $N=20$, the chaos transition occurs near
$p^* \sim 0.005$, corresponding to ${\sim}24$ surviving couplings out of
$\binom{20}{4} = 4{,}845$. The structural argument of
Sec.~\ref{sec:mechanism}---that $C(t)$ probes $H_\mathrm{int}$ while
sparsification acts on $H_{L,R}$---predicts the decoupling persists
through this transition, but directly probing the non-chaotic regime at
$N=20$ remains a target for future work, requiring either even sparser
systems ($p < 0.005$) or tensor network approaches. At $N=24$, the
level spacing analysis of Sec.~\ref{sec:chaos_transition}
(Table~\ref{tab:n24_level_spacing}) locates the chaos transition at
$p^* \approx 0.002$--$0.003$ (${\sim}21$--$33$ surviving couplings out
of $\binom{24}{4} = 10{,}626$). However, the full transmission signal
computation at $N=24$ ($d = 2^{24} \approx 1.7 \times 10^7$) was
computationally prohibitive: each matrix-vector product in the
$2^{24}$-dimensional coupled space requires two $4096 \times 4096$
dense matrix multiplications (${\sim}9$~seconds per matrix-vector product), and the Lanczos
basis ($m = 60$ Krylov vectors of dimension $2^{24}$) would require
${\sim}15$~GB, exceeding available memory. Computing the $N=24$
transmission signal through the now-identified chaos transition remains
a natural target for dedicated high-performance computing resources,
and would provide the first direct test of signal-chaos decoupling at
a system size where the non-chaotic regime retains a meaningful number
(${\sim}10$--$20$) of surviving couplings.

The dependence on the coupling structure is a more probing question. Our
$H_\mathrm{int} = i\mu\sum_j \psi_j^L \psi_j^R$ is bilinear and
site-diagonal: each left Majorana couples only to its right partner. A
non-diagonal coupling such as $H_\mathrm{int} \propto \sum_{ij}
M_{ij}\,\psi_i^L\psi_j^R$ with a random matrix $M_{ij}$ would entangle
the coupling structure with the internal dynamics, potentially breaking
the decoupling. If so, this would sharpen the structural explanation:
the decoupling holds precisely when $H_\mathrm{int}$ acts as a
``channel'' between the two sides without encoding information about the
internal Hamiltonian. Testing this would distinguish whether the
decoupling is a generic feature of bipartite coupled systems or a special
property of the standard wormhole protocol.

Our study is also restricted to the $q=4$ SYK model. The structural
argument of Sec.~\ref{sec:mechanism} does not depend on $q$, and we
expect the decoupling to hold for any even $q$ for which the dense
model is chaotic ($q \geq 4$). The $q=2$ SYK model is integrable at
all sparsities, so the chaos-to-integrable transition studied here does
not arise. Testing the decoupling at $q=6$ or $q=8$ would verify that
the result is not an artifact of the specific $q=4$ interaction
structure.

Similarly, we have studied only dephasing noise, which acts locally on
each qubit. Depolarizing noise, amplitude damping, and coherent gate
errors---the dominant noise sources on superconducting and trapped-ion
hardware---may couple to the internal dynamics differently. The noise
factorization $|C(t^*)| \approx f(\mu) \cdot g(\gamma)$ could break
for noise channels whose effect depends on the spectral structure of
$H_{L,R}$. Understanding which noise channels preserve the factorization
and which break it is directly relevant for experimental design: it
determines whether the gate complexity savings from sparsification come
with hidden noise costs.

Our evidence is purely numerical, and the absence of an analytical proof
is a significant gap. The variance rescaling argument provides an
intuitive explanation but does not constitute a derivation. In the
large-$N$ limit, the SYK model is solvable via Schwinger-Dyson equations
for the Green's function $G(\tau_1,\tau_2)$ and self-energy
$\Sigma(\tau_1,\tau_2)$~\cite{Maldacena:2016hyu}. The sparsified model
has the same saddle-point equations at leading order in $1/N$ (since the
variance rescaling preserves the relevant moments), which suggests the
transmission signal---computed from the two-point function across the
wormhole---should be identical at leading order. Making this argument
precise, including the identification of subleading corrections that
might distinguish dense from sparse, would place the decoupling on
rigorous footing. This observation also raises an important
interpretive caveat: if the large-$N$ saddle point is preserved by
variance rescaling, then the sparsified model retains the holographic
structure at leading order, and the signal invariance could be seen
as evidence that the sparsified model \emph{remains} holographic
rather than evidence that the signal is chaos-independent.
Distinguishing these two interpretations requires identifying
observables sensitive to subleading corrections where dense and sparse
models diverge---precisely the chaos diagnostics that we argue should
supplement the transmission signal in future experiments.

Perhaps the most important question our work raises is one of experimental
epistemology: \textit{what combination of observables would constitute
sufficient evidence for holographic dynamics in a quantum simulation?}
The transmission signal alone is clearly insufficient, as we have shown
it persists without chaos. Level spacing statistics alone are insufficient
because they diagnose chaos but not traversability. A natural proposal is
to require both: a transmission signal \textit{and} GUE-level statistics,
demonstrating that the system is both chaotic and exhibits inter-system
information transfer. But even this may not suffice, since Schuster et
al.~\cite{Schuster:2021uvg} showed that generic chaotic systems (not only
holographic ones) produce the same signal. Whether there exists any
finite-$N$ observable that distinguishes holographic from merely chaotic
dynamics remains an open and fundamental question for the quantum gravity
simulation program.

The distinction between ensemble and single-instance behavior also
deserves attention. Our invariance is a statement about ensemble
averages; individual disorder realizations at low $p$ have a standard
deviation of ${\sim}7\%$ of the mean, with worst-case excursions exceeding
$15\%$ (Fig.~\ref{fig:disorder}). Quantum hardware
experiments probe a single Hamiltonian instance, not an ensemble. At
$p \gtrsim 0.1$, where single-instance fluctuations are below ${\sim}3\%$,
the practical impact is modest. But at extreme sparsity, the ensemble
mean may be a poor predictor of any individual experiment. This
tension between ensemble-level theoretical predictions and
single-instance experimental reality is not unique to our
setting---it pervades the physics of disordered systems---but it is
particularly acute here because the experimental motivation relies on
specific quantitative features of the signal.

%=============================================================================
\section{Conclusion}
\label{sec:conclusion}
%=============================================================================

We have demonstrated numerically that the traversable wormhole transmission
signal in the doubled SYK model is controlled by the inter-system coupling
$\mu$ and is independent of the internal Hamiltonian complexity, as measured
by the chaos-to-integrable transition under sparsification. Five lines of
evidence support this conclusion. First, the ensemble-averaged peak height
at $N=10$ varies by less than 1.1\% as the level spacing ratio drops from
the GUE value ($\avg{r} \approx 0.60$) to deeply sub-Poisson statistics
($\avg{r} \approx 0.26$). Second, a 1{,}200-instance sweep over $\mu$
confirms that the signal depends strongly on the coupling but is
statistically independent of sparsity at each coupling value tested.
Third, the noise sensitivity of the signal is likewise independent of
sparsity, with the critical dephasing rate
$\gamma^* \approx 0.055\,J$ constant to within 1\% across all tested
sparsities. Fourth, the TFD state retains its thermal structure
(entanglement entropy within $0.2\%$, thermal fidelity at machine
precision) across the full sparsity range, even as the state vector
itself changes substantially. Fifth, a Krylov-subspace extension to
$N=20$ ($d \approx 10^6$, 150 instances) confirms signal invariance
to within $0.19\%$ over a $50\times$ sparsification range within the
chaotic regime accessible at this system size.
Additionally, a level spacing analysis at $N=24$ (10{,}626 total
couplings, 70 instances) locates the chaos transition at
$p^* \approx 0.002$--$0.003$, confirming that random-matrix
universality requires a roughly constant number of ${\sim}15$--$25$
coupling terms across three system sizes ($N=10$, $14$, $24$) and
establishing the precise sparsity threshold for future signal
studies at larger $N$.
These findings indicate that the transmission signal is a
diagnostic of the inter-system coupling fidelity rather than a probe of
quantum chaos. On the interpretive side, this means that future quantum
gravity simulation experiments should supplement the transmission signal
with independent chaos diagnostics to substantiate claims of holographic
behavior. On the practical side, the invariance implies that the vast
majority of the Hamiltonian's coupling terms can be discarded without
affecting the signal, reducing the quantum gate count per Trotter step by
a factor of ${\sim}50$ at $N=10$ and by substantially larger factors at
experimentally relevant system sizes---bringing traversable wormhole
simulations within reach of current quantum hardware.
All numerical data, analysis code, and figure-generation scripts are
available at~\url{https://github.com/sagardubey473/syk-wormhole-signal}.

\bibliography{references}

\end{document}